%% file: Main.tex
\documentclass[conference]{IEEEtran}
\IEEEoverridecommandlockouts
\usepackage{cite}
\usepackage{amsmath,amssymb,amsfonts}
\usepackage{algorithmic}
\usepackage{graphicx}
\usepackage{textcomp}
\usepackage{xcolor}
\usepackage{subcaption}
\usepackage{array,tabularx}
\usepackage{tcolorbox}

\usepackage{lipsum}
\def\BibTeX{{\rm B\kern-.05em{\sc i\kern-.025em b}\kern-.08em
    T\kern-.1667em\lower.7ex\hbox{E}\kern-.125emX}}
\begin{document}
\newenvironment{conditions*}
  {\par\vspace{\abovedisplayskip}\noindent
   \tabularx{\columnwidth}{>{$}l<{$} @{${}={}$} >{\raggedright\arraybackslash}X}}
  {\endtabularx\par\vspace{\belowdisplayskip}}
  
\newcommand*{\todo}{\textcolor{red}}
\newcommand*{\done}{\textcolor{green}}
\newcommand*{\notetext}{\textcolor{blue}}
  
\title{Monte Carlo Methods for Industry~4.0 Applications
}

\author{\IEEEauthorblockN{Petr Kostka}
\IEEEauthorblockA{\textit{Masaryk University}\\
Brno, Czech Republic \\
500331@mail.muni.cz}
\and
\IEEEauthorblockN{Bruno Rossi}
\IEEEauthorblockA{\textit{Masaryk University}\\
Brno, Czech Republic \\
brossi@mail.muni.cz}
\and
\IEEEauthorblockN{Mouzhi Ge}
\IEEEauthorblockA{\textit{Deggendorf Institute of Technology} \\
Deggendorf, Germany \\
mouzhi.ge@th-deg.de}

}

\maketitle

\begin{abstract}
The fourth industrial revolution and the digital transformation, commonly known as Industry 4.0, is exponentially progressing in recent years. Connected computers, devices, and intelligent machines communicate with each other and interact with the environment to require only a minimum of human intervention. An important issue in Industry 4.0 is the evaluation of the quality of the process in terms of KPIs. Monte Carlo simulations can play an important role to improve the estimations. However, there is still a lack of clear workflow to conduct the Monte Carlo simulations for selecting different Monte Carlo methods. This paper, therefore, proposes a simulation flow for conducting Monte Carlo methods comparison in Industry~4.0 applications. Based on the simulation flow, we compare Cumulative Monte Carlo and Markov Chain Monte Carlo methods. The experimental results show the way to use the Monte Carlo methods in Industry~4.0 and possible limitations of the two simulation methods.
\end{abstract}


\section{Introduction}

Industry 4.0 is a strategic approach that is used to digitize processes in manufacturing~\cite{erboz2017define}, creating automated industries through human-machine connections replacing human workers in dangerous or unsuitable tasks to ease the working environment and create new, more suitable positions~\cite{erboz2017define, vaidya2018industry, thoben2017industrie}. One of the critical features of Industry 4.0 is the creation of highly automated industries through human-machine connections. Creating more digitized systems and network integration via smart systems will replace human workers in specific tasks to ease the working environment~\cite{erboz2017define,vaidya2018industry}. 

Applications in Industry 4.0 aim at operational efficiencies and accelerated growth in productivity which leads to corporate profitability improvement. It leads to overall agility and flexibility in production due to increasing individual demand and will enable cost reduction manufacturing and better risk management. Integrated processes and human-machine interaction influence the industry to operational efficiency in productivity, cost and time~\cite{erboz2017define}.


Since Industry 4.0 leads to the generation of large amounts of data that needs to be analyzed, one of the main issues if the system is able to predict the quality of the process in terms of Key Performance Indicators (KPIs). For this reason, simulations are one of the main pillars of Industry 4.0~\cite{ghobakhloo2020industry}. However, it is still unclear regarding how to select a proper Monte Carlo simulation for Industry 4.0 applications. An incorrect selection of Monte Carlo simulation may lead to the lower performance of the application or even inaccurate results. Therefore, Monte Carlo model selection is critical for Industry~4.0 applications. 

This paper therefore proposes a simulation flow to conduct Monte Carlo method comparison for Industry~4.0 applications. Based on the workflow, we intend to perform the model selection with two widely-used Monte Carlo methods, which are \emph{Cumulative Monte Carlo Method} (MC) and  \emph{Markov Chain Monte Carlo Method} (MCMC). The Monte Carlo model selection can bring many benefits such as increased efficiency and productivity~\cite{tortorella2018implementation,hedvivcakova2019benefits}. Furthermore, in this work we will use MC simulations to analyze real-world data, calculate the KPIs and then capture them in the probability distribution. The data can also be used for later prediction in the industrial context.

The rest of the paper is structured as follows. Section II describes the background of work by focusing on the pillars of Industry 4.0. Section III proposes the workflow for Monte Carlo simulation and compares two Monte Carlo methods, Section IV conducts an experiment to evaluate the advantages and disadvantages of the two methods, and proposes the limitations as well as recommendations of usage of the Monte Carlo  methods under which circumstances. Finally, Section V concludes the paper and outlines future research works.

\section{Background}


Industry 4.0 aims at operational efficiencies and accelerated productivity growth, leading to overall agility and flexibility in production due to increasing individual demand. Also, it will enable cost reduction manufacturing and better risk management to improve the manufacturer's profitability. Nevertheless, the impacts of Industry 4.0 are far more extensive as they also cover waste reduction, harmful gas emission reduction and much more~\cite{ghobakhloo2020industry}. Specific limitations may represent the complexity of the whole problem associated with Industry 4.0 in the form of vast financial expenses or with the presence of cyber security risks~\cite{vaidya2018industry}. When considering some of the nine pillars of Industry 4.0, like simulations, IIoT or cyber security, companies are transformed into smart factories, which symbolizes a fully interconnected manufacturing system for performing tasks using data to produce goods. Below we discuss each area associated to Industry 4.0~\cite{ghobakhloo2020industry,vaidya2018industry,erboz2017define}.

\begin{itemize}
    \item \textbf{Big Data and Analytics} - This concept refers to the collection and evaluation of data from various production sources and, together with the customer management system, serves for real-time decision making of their strategy. At the same time, it is also used to search and identify errors in data records.
    \item \textbf{Autonomous Robots} - It refers to the interaction of autonomous robots during the execution of tasks together with humans and at the same time learning from them. Robots should also execute very complex or otherwise difficult functions for a human worker.
    \item \textbf{Simulation} - Use real-time data to simulate the real world in a virtual model to improve many aspects, including decision making and planning operations, using information and accurate estimates obtained from the simulation.
    \item \textbf{Horizontal and Vertical System Integration} - Horizontal integration is used to integrate partners within supply chains, and vertical integration refers to flexible systems within the factory, both integrated to achieve agility.
    \item \textbf{The Industrial Internet of Things} - IIoT represents a network of interconnected devices that communicate via standard protocols and gather data. From collected data, computers make the decision about operations.
    \item \textbf{The Cloud} - According to the concept of Industry 4.0, it is critical for the connectivity and communication between elements, where organizations require increased data sharing between manufacturing sites and companies.
    \item \textbf{Cyber security} - For increased connectivity and communication, it is critical to protect crucial industrial systems with the help of preventable solutions and defence systems; otherwise, cyber-attacks could have destructive effects on the business model.
    \item \textbf{Additive Manufacturing} - Refers to producing customized goods for the requirements of customers methods such as 3D printing enables to create goods on customers demand and not to be fully stocked.
    \item \textbf{Augmented reality} - Is the interactive technology which allows the virtual world to be part of the real world with needed information and surroundings. This allows human-machine interaction to be used in production.
\end{itemize}

Improving productivity is a key for being competitive in a global market~\cite{fleischer2006calculation}. KPIs are quantifiable values for measuring and improving the manufacturing process. These metrics show the level of performance of energy, material, time, etc. With the increasing usage of sensors with real-time data, it is easier for manufacturers to examine data using KPI indicators~\cite{brundage2017using}. The most used KPI in manufacturing is Overall Equipment Efficiency (OEE), which reflects three separate and measurable components: Availability, Performance and Quality~\cite{iannone2013managing}. Another useful KPI, Efficiency, shows the percentage of completion from a current target in a given interval of time. Moreover, with the wider adoption of Industry~4.0, KPIs are gaining importance in measuring and mainly improving these indicators by using simulation methods~\cite{ghobakhloo2020industry,guban2017industry}.

\section{Monte Carlo Simulation}
\label{sec:MCS}


\subsection{Cumulative Monte Carlo Method} \label{subsec:mc}
The stochastic method called MC simulation uses Pseudo-Random Number Generators (PRNG) for random (independent) sampling and statistical analysis to estimate parameters of mathematical functions, imitating the operations of complex systems~\cite{harrison2010introduction}. For example, in the concept of Industry~4.0, MC simulation can be used for estimating the success of cyber-attacks~\cite{hanischestimating}, creating a maturity model and tools to assist manufacturing companies in evaluating the progress of Industry~4.0 initiatives and digitalization. The MC method is closely related to examining random experiments, where the result cannot be determined. The main idea behind this simulation method can be interpreted simply as finding the expected value resulting from repetition of a random event. This method is used in practically every field such as science, finance, risk analysis, and engineering.

The key idea of this method is to take a sufficient amount of obtained data, use the cumulative frequency of the data points to define its probability distribution and fitting the Cumulative Distribution Function (CDF). The next step is to randomly select values from the obtained cumulative distribution and thus get the resulting simulated data. This basic MC simulation is easy to implement, and its main advantage over other more sophisticated methods is the speed and especially the independence of probabilistic data distributions. The following procedure is typically applied~\cite{molak1996fundamentals}:


        

\begin{enumerate}
    \item Minimum and maximum for the empirical distribution (usually outside of the range of the observed data) are added into data based on knowledge of the variable.
    \item Data points are sorted in ascending order.
    \item Cumulative probability \textit{P(x)} is calulated as follows:
        \begin{equation}
            P(x_i) = \frac{i}{n+1}
        \end{equation}
        where:
        \begin{conditions*}
            P(x_i)    &  cumulative probability of the $i^{th}$ data point\\
            n     &  total size of elements \\
        \end{conditions*}
\end{enumerate}

\subsection{Markov Chain Monte Carlo Method} \label{subsec:mcmc}
Another sampling technique from a probability distribution is called Markov Chain Monte Carlo (MCMC), which uses Markov Chains for the process of samples creation. The main difference between MC and MCMC methods is that samples in the MC method are independent. In contrast, in the MCMC method, the next sample is dependent on the last sample that was approved. With MCMC, we can sample from a distribution we cannot compute or identify. MCMC has been successfully used in many areas - for example, multi-dimensional integral, computational physics, numerical approximations, Bayesian statistics, or to compute parameters of probabilistic distributions~\cite{rubinstein2016simulation,von2011bayesian}. 

For using MCMC algorithms, it is important to understand principles of Bayesian statistics, which is statistical theory dependent on the interpretation of Bayesian probability~\cite{von2011bayesian}. The Bayes theorem describes the conditional probability called posterior. The specific case is Bayesian inference which is part of statistical inference, where the Bayes theorem is used to gain knowledge about the distribution~\cite{bolstad2016introduction}. More specifically, given the data $X$ with parameters $\theta$ we are able to get the posterior distribution $f(\theta | X)$.

\begin{equation}
f(\theta|X) = \frac{f(X|\theta)f(\theta)}{f(X)}
\end{equation}
where:
        \begin{conditions*}
            f(X)    &  normalizing constant\\
            f(\theta)     &  prior distribution \\
            f(X|\theta))     &  likelihood distribution \\
            f(\theta|X)     &  posterior distribution \\
        \end{conditions*}

\bigskip
There are several MCMC algorithms, which differ in the way of creating the Markov Chain, but the general idea remains the same. Samples are taken from the constructed Markov Chain by the Monte Carlo method from the probability distribution. The target is to wait until the chain has converged to find an optimal value. The Metropolis-Hasting algorithm is usually used to draw samples whose distribution converges to the posterior distribution~\cite{REIS200597}.
The sample is then examined by an acceptance probability, which checks if the sample can be added to a chain or not. This algorithm is far more flexible than the others~\cite{rubinstein2016simulation}. The Burn-in (Warm-up) method is used to optimize the parameters. The method discards a certain amount of initial samples of the Markov Chain not to distort the final result since the values are in more confident areas~\cite{bolstad2016introduction}. Secondly, it removes the initial values of our estimate because they are usually inaccurate as they were set manually. The recommended amount is usually about 1\% or 2\%, since higher values would be inefficient~\cite{geyer1992practical}.


The goal of MCMC simulation is to take the collected data~$d$, given model parameters $\theta$ and prior knowledge (belief) to estimate the distribution. In the Bayesian context, the joint posterior distribution is the key of interest, and the Bayesian rule gives the model~\cite{mcmc_mh}:

\begin{equation}
P(\theta | d) \propto P(d | \theta)P(\theta)
\end{equation}
where:
\begin{conditions*}
P(\theta | d)    &  joint posterior distribution\\
P(d | \theta)     &  likelihood function for model parameters \\
P(\theta) &  prior distribution
\end{conditions*}

Since posterior distribution $P(\theta | d)$ cannot be computed analytically, it is possible to calculate the value of a likelihood function $P(d | \theta)$ for a given parameter realization. An abstract Markov Chain is created with states of $\theta$ parameter values so that the stationary distribution of this chain is the result of the posterior distribution. For the proposal distribution, it is usually recommended to use normal distribution~\cite{haario2005componentwise}. For the construction of the Markov chain, if proposal distribution is symmetric, the convention is to apply Metropolis-Hastings  algorithm~\cite{mcmc_mh, strens2003evolutionary}:

\begin{enumerate}
    \item Obtain $\theta^{(0)}$ - the initial realization of the parameter vector
    \item Obtain $\theta^{(n)}$ - the $n^{\text{th}}$ realization of the parameter vector by:
    \begin{enumerate}
        \item Sampling from proposal distribution to get current candidate parameter vector $\theta^*$
        \item Calculate an acceptance probability $\alpha$:
        \begin{equation}
        \alpha = \frac{P(\theta^* | d)}{P(\theta^{(n-1)}) | d} = 
        \frac{P(d | \theta^*) P(\theta^*)}{P(d | \theta^{(n-1)}) P(\theta^{(n-1)})}
        \end{equation}
        \item Get $u$ from Uniform[0,1], if u $<$ min(1, $\alpha$), set $\theta^n = \theta^*$, otherwise set $\theta^n = \theta^{(n-1)}$
    \end{enumerate}
\end{enumerate}

\begin{figure}[!h]
    \begin{subfigure}[b]{.49\linewidth}
    \includegraphics[width=\linewidth]{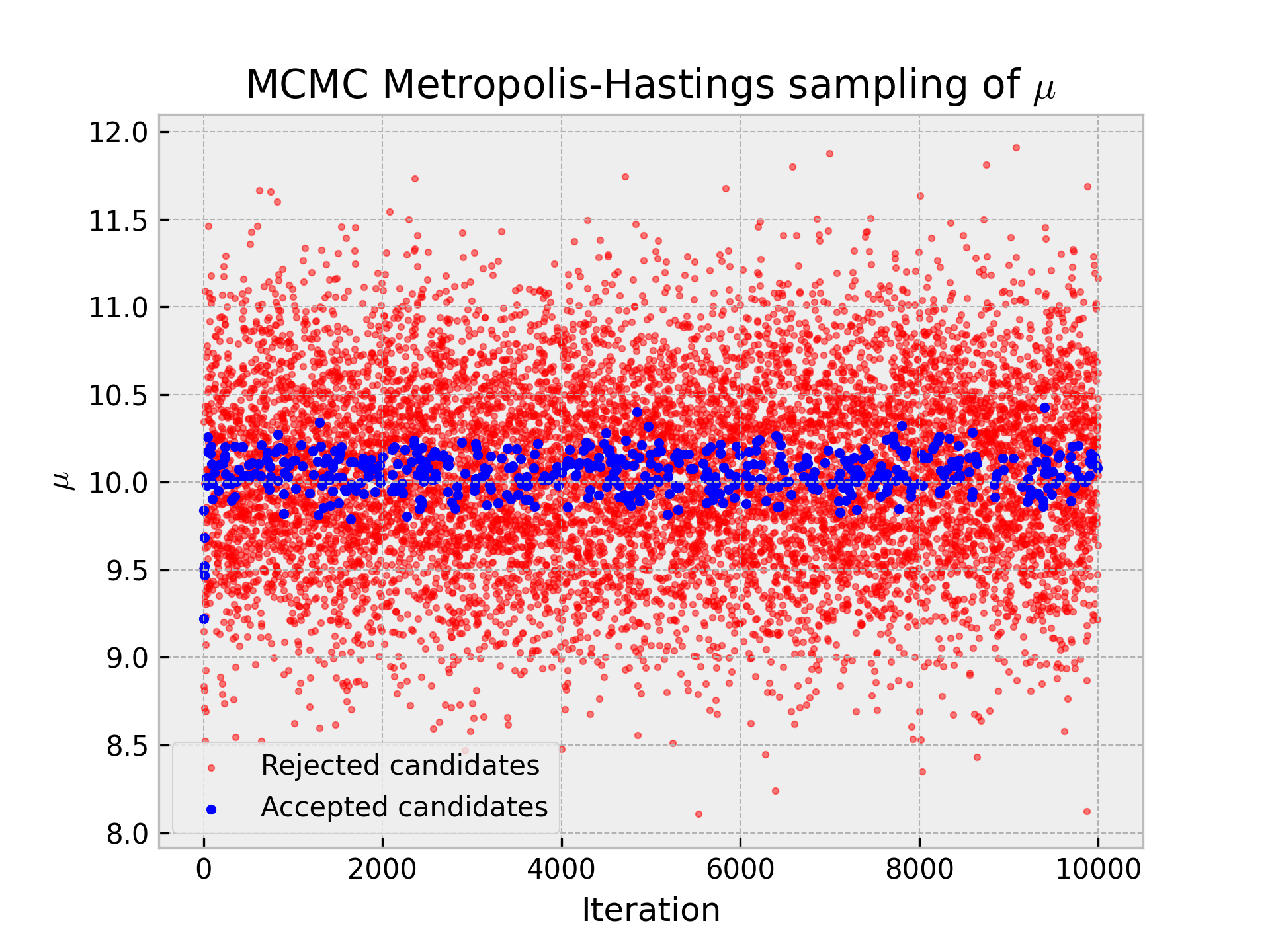}
    \end{subfigure}
    \begin{subfigure}[b]{.49\linewidth}
    \includegraphics[width=\linewidth]{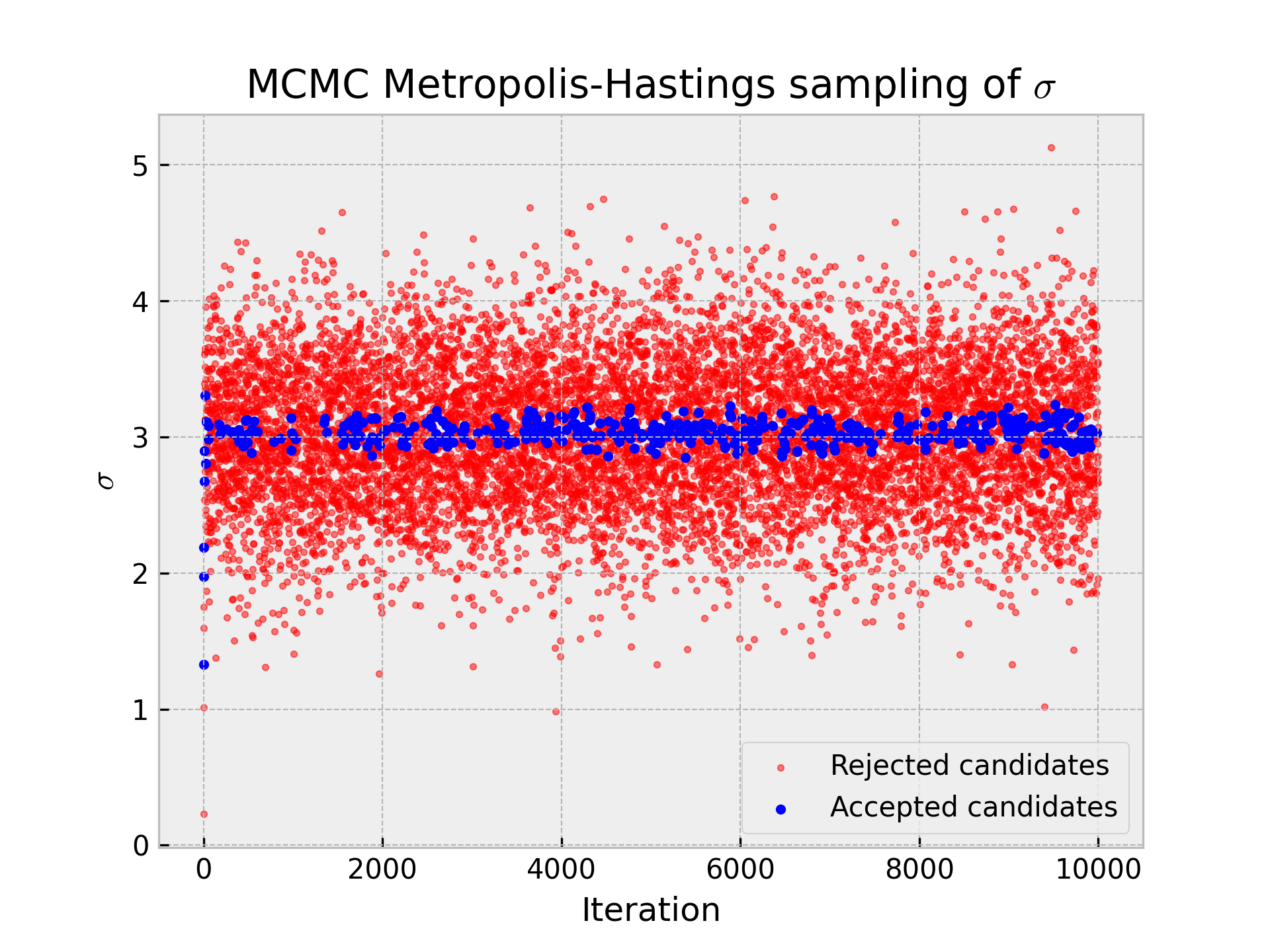}
    \end{subfigure}
    \caption{MCMC Metropolis-Hastings sampling for parameters $\mu$ and $\sigma$}
    \label{fig:MCMC_MH_sampling}
\end{figure}

\bigskip
The generated Markov chain will eventually (after a sufficient number of iterations) converge to the joint posterior distribution for any form of proposal distribution~\cite{gilks1995markov}. The convergence of the solution for parameters $\mu$ and $\sigma$ is shown on Figure~\ref{fig:MCMC_MH_sampling}, where samples were taken from normal distribution with parameters $\mu$=10 and $\sigma$=3.

Since implementation of a solution for the given dataset is following a normal probability distribution, this paper will present the calculation of the probability acceptance $\alpha$ for this case. There are two parameters for the normal distribution - $\theta$ consisting of $\mu$ and $\sigma$. In addition, it is recommended to calculate log-likelihood instead of likelihood. It helps numerical stability because the multiplication of thousands of small values can cause underflow in the system's memory~\cite{chandra2011robust, taboga2017lectures}. The logarithm is a solution because it modifies a product of densities into a sum. From these findings, we present here the expression of the final form for the calculation of acceptance probability~$\alpha$:

\begin{equation}
  \begin{aligned}
    \alpha & = \frac{P(d | \theta^*) P(\theta^*)}{P(d | \theta^{n-1}) P(\theta^{n-1})}\\
      & =  \frac{P(d | \mu^*,\sigma^* ) P(\mu^* , \sigma^*)}{P(d | \mu^{n-1} , \sigma^{n-1}) P(\mu^{n-1} , \sigma^{n-1})}\\
  \end{aligned}
\end{equation}

\begin{equation}
  \begin{aligned}
    ln(\alpha) & = ln(P(d | \mu^*,\sigma^* )) + ln(P(\mu^* , \sigma^*))\\
      &\quad - ln(P(d | \mu^{n-1} , \sigma^{n-1})) + ln(P(\mu^{n-1} , \sigma^{n-1}))
  \end{aligned}
\end{equation}

where:
\begin{equation}
ln(P(d | \mu,\sigma)) = - \frac{n}{2} ln(2\pi) - \frac{n}{2} ln(2\sigma^2) - \frac{1}{2\sigma^2} \sum_{i=1}^{n} (d_i - \mu)^2 
\end{equation}

The prior distribution summarizes our beliefs about the value of $\mu$ and $\sigma$ before we observe any data. As we do not know any information (without data) about the distribution, we can apply only basic knowledge for Gaussian distribution, that is $\sigma$~$>$~0.


\section{Experimental Analysis}

The experiment compares MC and MCMC methods with two different Industry 4.0 value streams M1, M2 using \textit{Efficiency} as the main KPI for testing the methods. More specifically, we first create four scenarios - two using simulated data and two using real-world data from value streams M1, M2. 100~stochastic datasets are created for each scenario. Then we compute the confidence interval at 90\% confidence level for each stochastic dataset and compute coverage probability.


As it is challenging to determine the accuracy of the result regarding real-world data objectively, the method presented in~\cite{comparison} compares uncertainty methods using probability coverage when the true value is occupied in the confidence interval. However, the authors of this method compare the data generated by the mathematical model and add observation noise to this data to add realism. Based on this method, we will proceed to compare implemented MC and MCMC, where we combine simulated data with added observation noise and real-world data obtained directly from the manufacturer. 


\subsection{Method}
\begin{figure}[htb]
        \centering
            \includegraphics[width=0.95\linewidth]{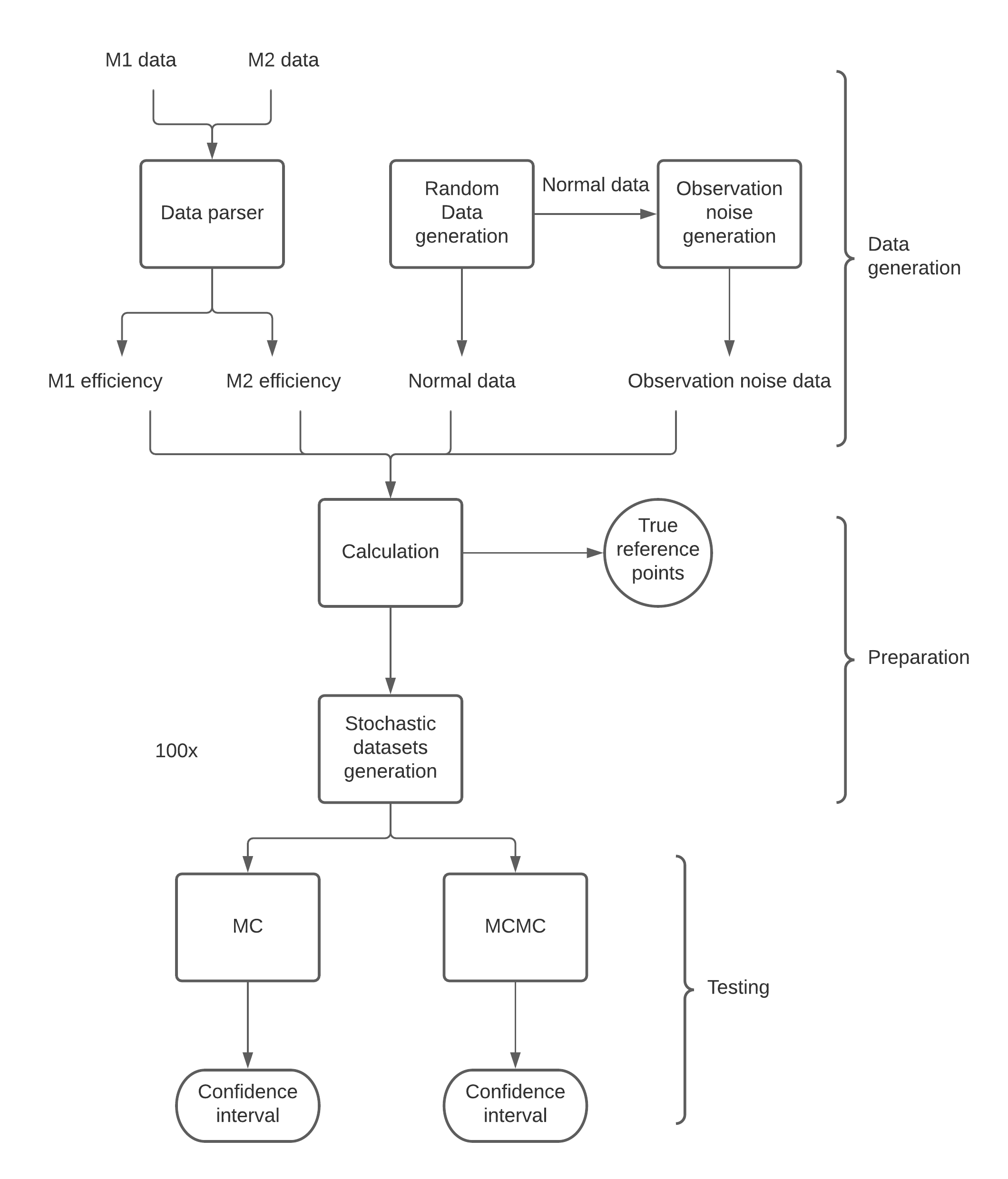}
            \caption{Workflow for the simulation procedure comparison}
            \label{simulation_procedure_comparison}
\end{figure}

The occurrence of the true value inside the generated confidence interval, also called \textit{coverage probability}, from the dataset will be the scale of each method performance for different datasets. The optimal value of coverage probability should be approximately equal to the confidence level. In Figure~\ref{simulation_procedure_comparison} we present  a list of the necessary phases of the workflow with a description for performing the simulation.

\subsubsection{\textbf{Data generation phase}}

\begin{itemize}
    \item \textbf{Data parser} - Collected data from the manufacturer come with two different datasets M1 and M2. Both datasets are divided into working shifts by twelve hours. Therefore every shift has the needed values for calculating KPIs per shift. We chose the KPI \textit{Efficiency}, as it is one of the main KPIs for the manufacturer. Among other things, this data follows a skewed normal distribution. Efficiency represents the actual quantity of parts produced as a percentage of the target quantity to be produced. For example, if sensors measured the actual quantity 120 parts and the target was 120 parts for this shift, the efficiency was one (or 100\%). Therefore if the value is less than one, it symbolizes insufficient results. If the value is equal to one or even better, it is greater than one, it means that shift was very efficient in terms of parts produced.
    
    \begin{equation}
    Efficiency =  \frac{Actual\: quantity\: produced}{Target\: quantity\: to\: be\: produced}
    \end{equation}
    
    \item \textbf{Random data generation} - Python library NumPy using function \texttt{numpy.random.normal}, which draws random samples from a normal (Gaussian) distribution, is used as a reference to compare ideally distributed data with no observation noise. 
    \item \textbf{Observation noise generation} - Some observation noise will be added to the original values from normal data in the interval [{-0.2,0.2}].
\end{itemize}

\subsubsection{\textbf{Preparation phase}}
\begin{itemize}
    \item \textbf{Calculation} - With given input data, two types of reference points are evaluated as the potential value of our interest. $\mu$ (mean) and $\sigma$ (standard deviation) can be used as a general measure for all kinds of data distribution. The true value of reference points are calculated 
    and will be used as true reference points for fitting in confidence intervals.
    
    \item \textbf{Stochastic datasets generation} - As datasets are based on four different scenarios, we will generate 100 stochastic datasets for every one of them, consisting of 100 observations sampled from the original dataset. For a normal data scenario, we will generate an extra 100 stochastic datasets with 1500 observations sampled from the original dataset to show both methods.
\end{itemize}

\begin{figure}[htb]
    \begin{subfigure}[b]{.49\linewidth}
    \includegraphics[width=\linewidth]{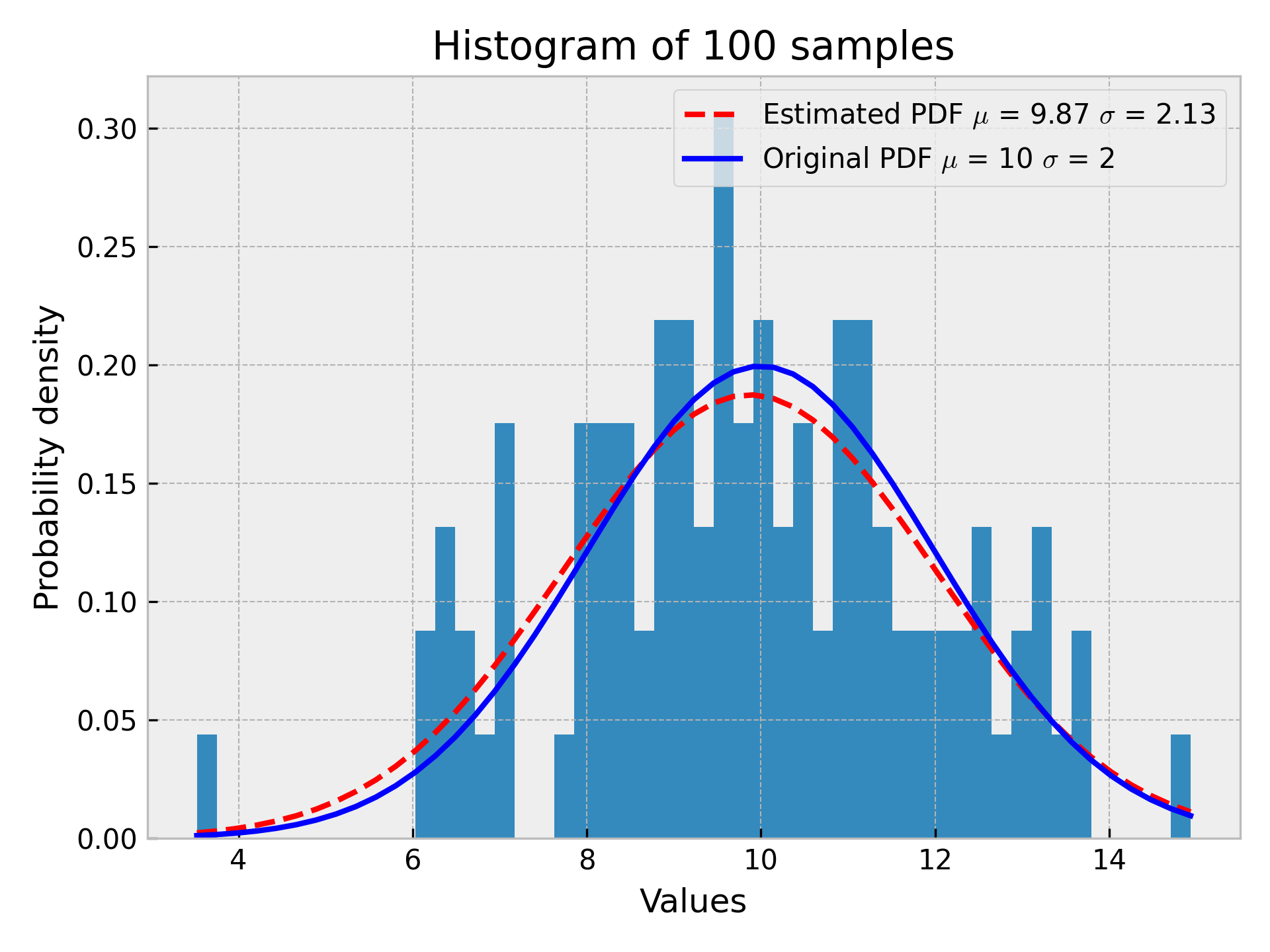}
    \end{subfigure}
    \begin{subfigure}[b]{.49\linewidth}
    \includegraphics[width=\linewidth]{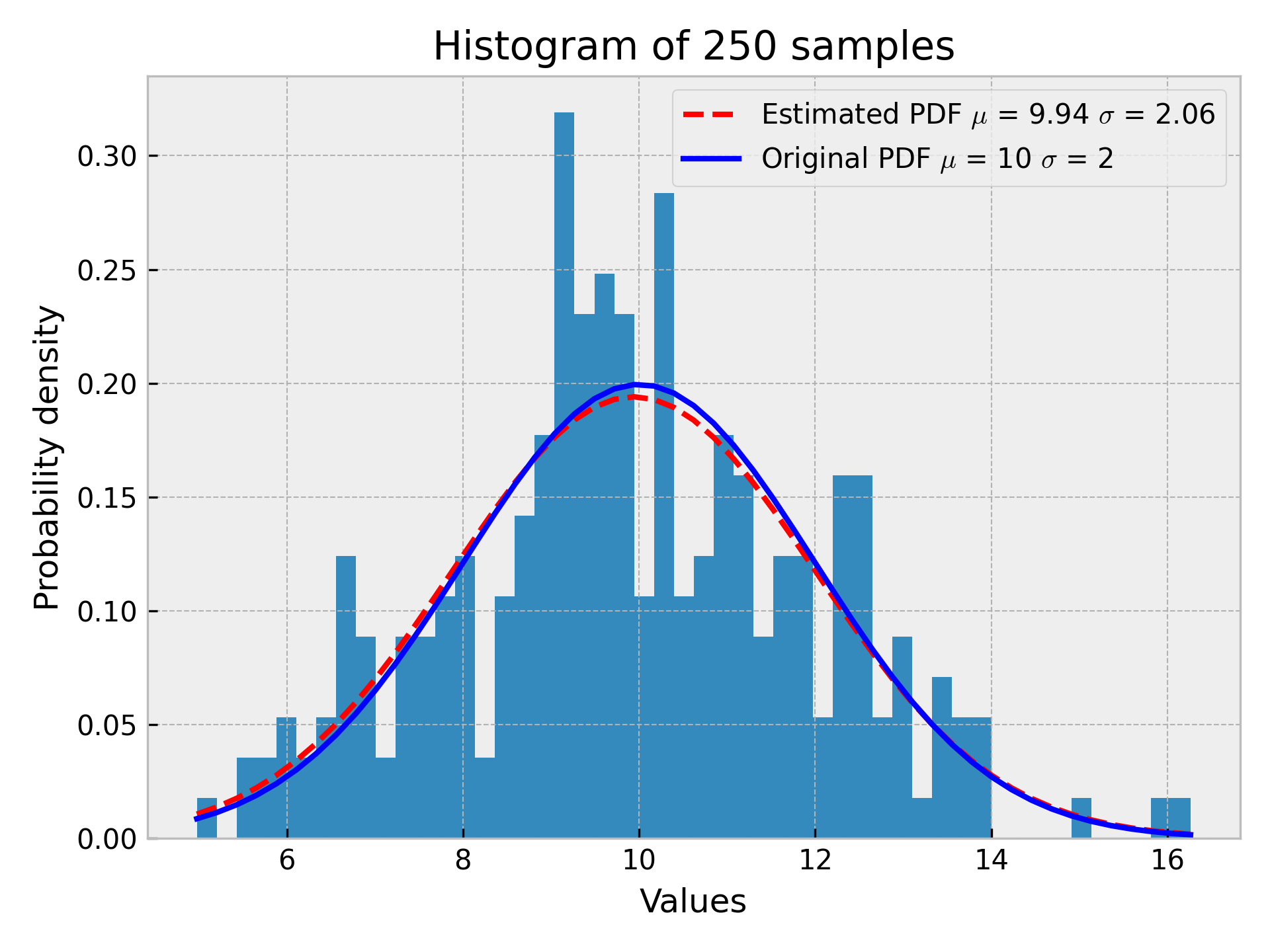}
    \end{subfigure}
    
    \begin{subfigure}[b]{.49\linewidth}
    \includegraphics[width=\linewidth]{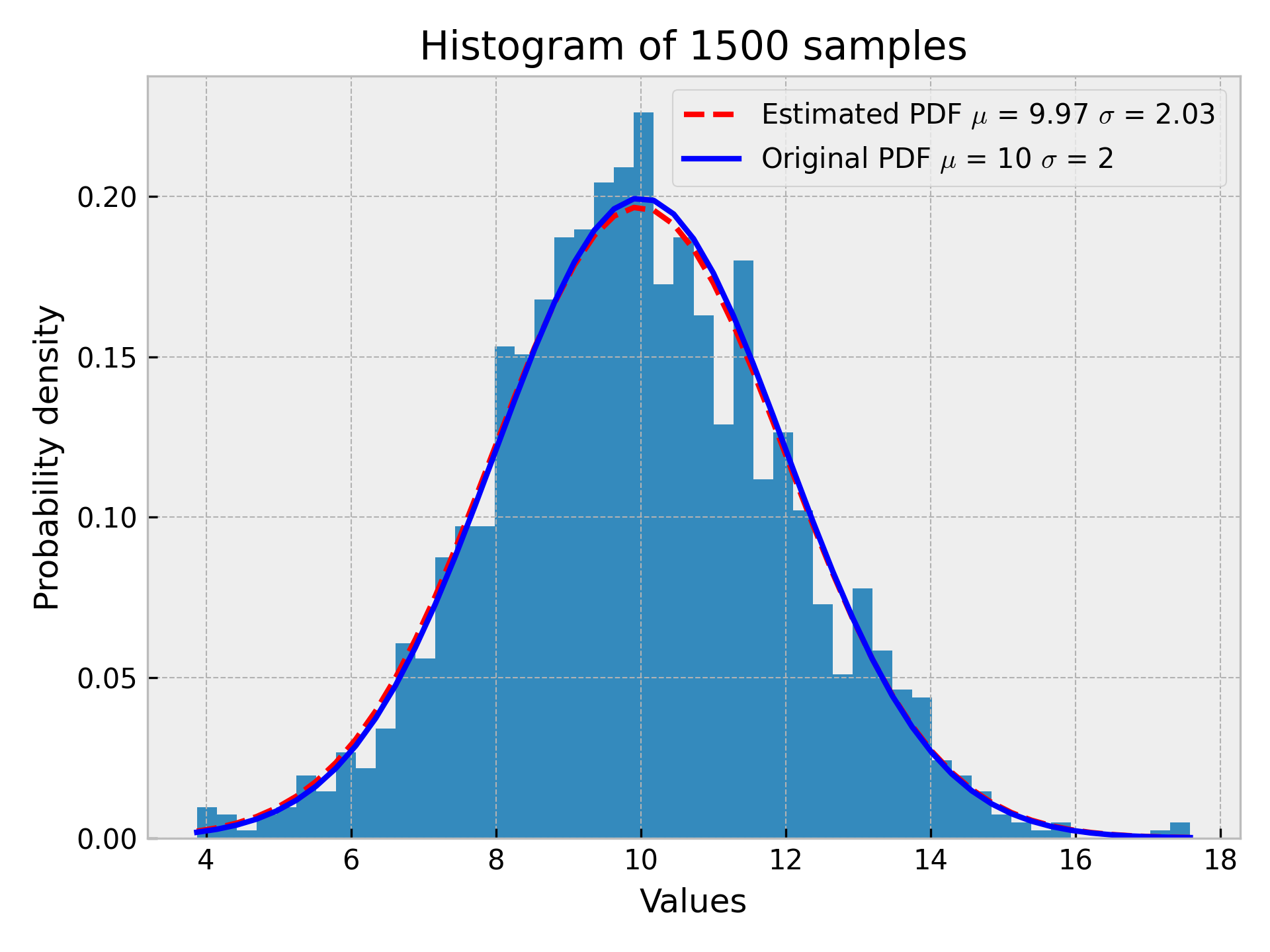}
    \end{subfigure}
    \begin{subfigure}[b]{.49\linewidth}
    \includegraphics[width=\linewidth]{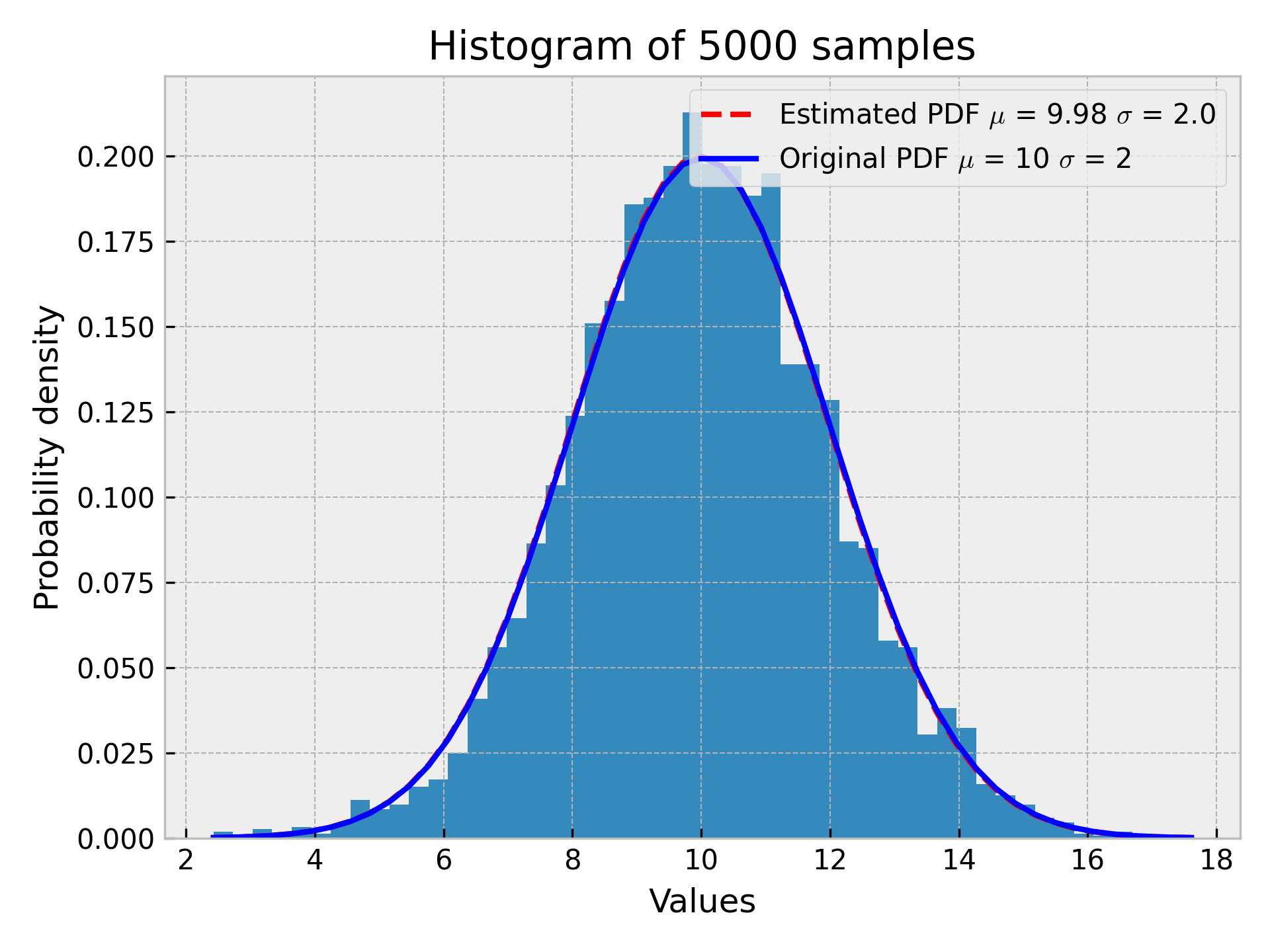}
    \end{subfigure}
    \caption{Convergence of MCMC method for a different count of collected samples}
    \label{fig:Comparison_CI}
\end{figure}

\begin{figure*}[htb]
\centering
    \begin{subfigure}[b]{.38\linewidth}
    \includegraphics[width=\linewidth]{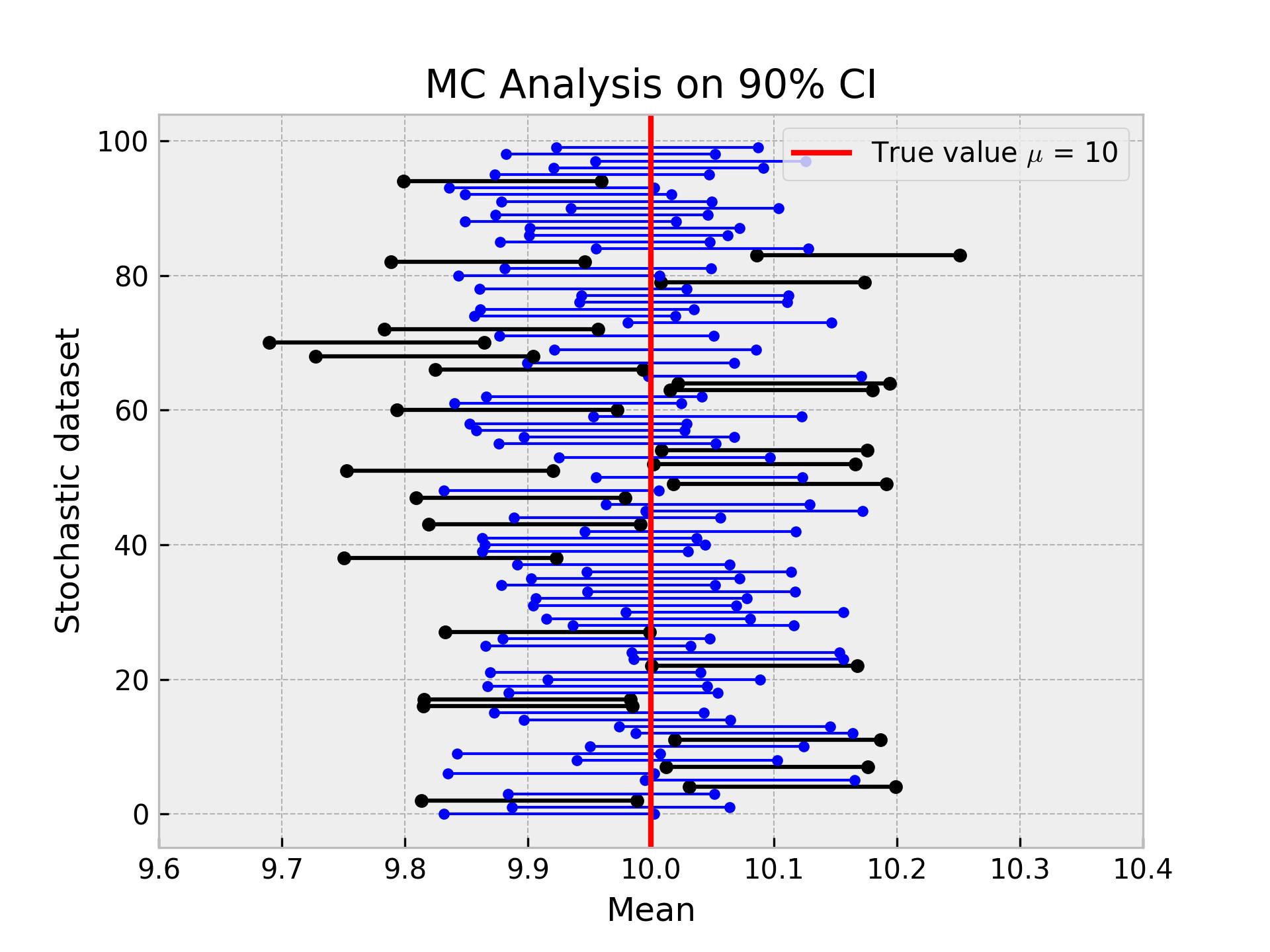}
    \end{subfigure}
    \begin{subfigure}[b]{.38\linewidth}
    \includegraphics[width=\linewidth]{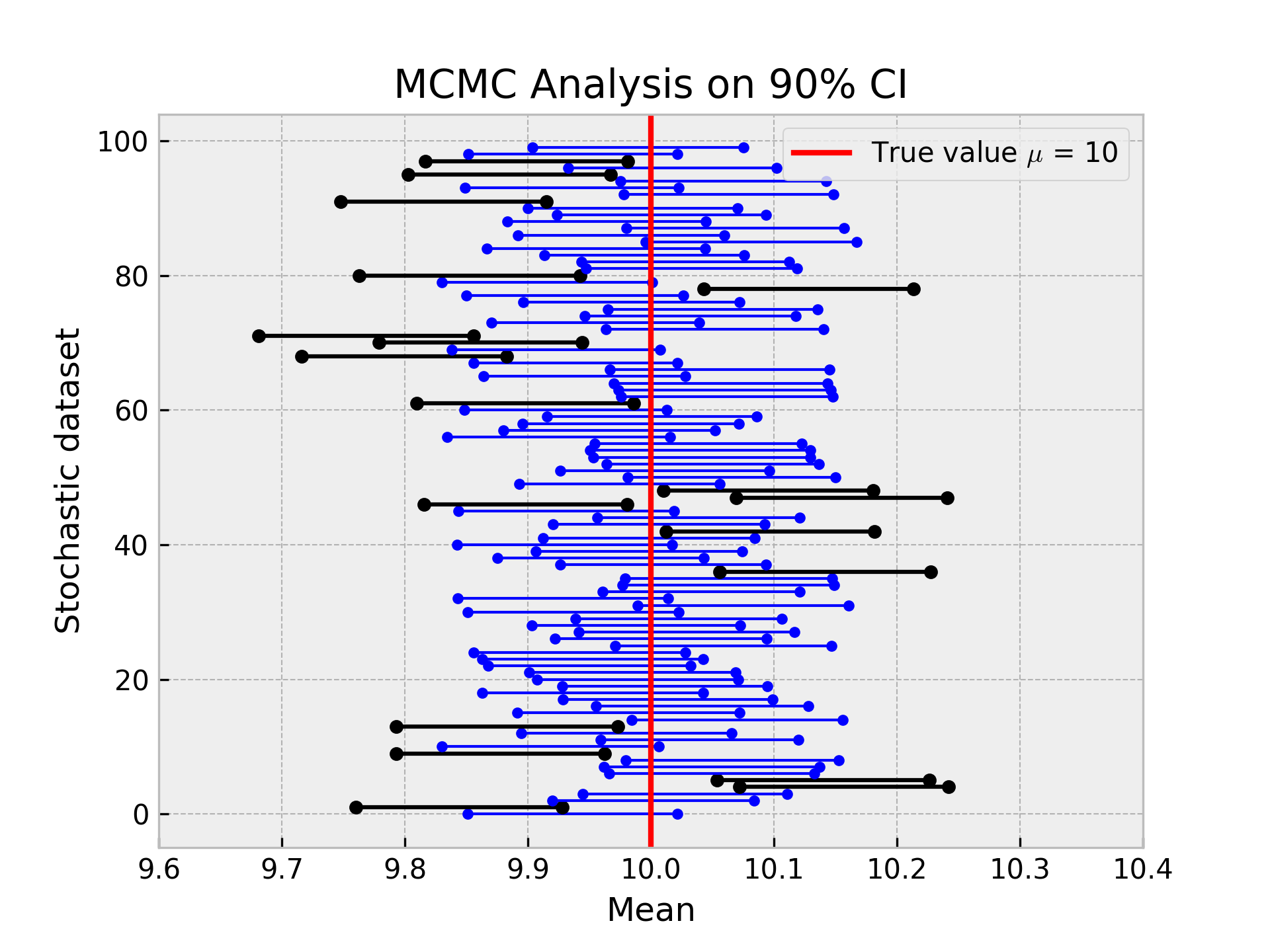}
    \end{subfigure}
    \caption{Confidence interval comparison on 100 stochastic datasets}
    \label{fig:MCMC_Convergence}
\end{figure*}

\subsubsection{\textbf{Testing phase}}
\begin{itemize}
    \item \textbf{MC} - The method designed in Section~\ref{subsec:mc} takes a stochastic dataset and, as an output, gives a CDF. 
    
    \item \textbf{MCMC} - The uncertainty method designed in Section~\ref{subsec:mcmc} takes a stochastic dataset and, as an output, gives probability distribution.
    
    \item \textbf{Confidence interval} - For every probability distribution created by method MC or MCMC, it is required to construct a confidence interval so that we would be able to compute coverage probability. Since probability distribution produced by MC is created by cumulative frequency, it does not follow the exact type of continuous probability distribution such as Gaussian, Gamma and others. For this case, it is recommended to use MC bootstrapping for computing confidence interval value~\cite{diciccio1996bootstrap}. To remain consistent, we will use this approach for every uncertainty method. For best performance, we used the already implemented bootstrapping method with bias correction with the Python library SciPy by means of \texttt{scipy.stats.bootstrap}.

\end{itemize}

\subsection{Results}

To test the performance of the confidence interval in both methods, we use the value of the confidence level of 90\% according to the recommendations from the literature in Section~\ref{sec:MCS}, which also states that the results are proportional to the different levels of confidence intervals.

Before starting the comparison, Figure~\ref{fig:MCMC_Convergence} demonstrates the dependence of the estimate's accuracy on the number of samples collected. It can be seen that under 100 samples, the accuracy would not be sufficient, but around 1500 samples, the method appears to be very accurate if the data follow the given probability distribution.

The calculated confidence intervals for 100 stochastic datasets from the normal data are shown in Figure~\ref{fig:Comparison_CI}. There is a comparison of the MC result with the MCMC, where the confidence intervals that contain the true value $\mu=10$ are shown in blue colour and those that do not include this value are shown in thicker black lines. The size of each stochastic dataset was 1500 and it is to demonstrate the ability of each method with the most significant knowledge. The left graph shows the MC analysis, where the probability coverage is 74 (26 intervals do not contain the true value). The MCMC method performed better, where the coverage probability reached the value of 81 (19 intervals do not include the true value) and is very close to the nominal value of 90.

From two methods, two reference values, four scenarios, 100 stochastic datasets, a total of 2000 confidence intervals were generated, from which the coverage probability was calculated in Table~\ref{tab:Comparison_CP}. Looking at all the data, none of the methods exceeded the probability coverage of 90, so there is a trend that the analysis will cover 90\% confidence interval less than 90\% time - this trend occurs for any chosen confidence interval. Since the data from the manufacturer is in the range of 100 to 150, to bring these simulations closer to these real-world values, stochastic datasets have the size of 150 for the generated normal data. There is a trend that MCMC performs better if there is enough data available. However, if the results are averaged over two reference points, then both methods show similar results.

\subsubsection{\textbf{Limitations}}
\begin{itemize}
    \item The smaller sample size of stochastic datasets is the leading cause of the lower coverage probability as it is shown on Table~\ref{tab:Comparison_CP}.
    \item Inaccuracies must considered for the result from real M1 and M2 data, as the true value is calculated from these samples and not from the probability distribution as in the other data. This can lead to optimistic results.
    \item For MCMC, the main limitation is the need to know the probability distribution to calculate the distribution parameters. In comparison, the cumulative MC method is generic for any type of data distribution.
    \item In terms of time and computational complexity, MCMC is more than 50 times slower than MC.
\end{itemize}

\begin{table*}[htb]
    \centering
    \caption{\label{tab:Comparison_CP}Table of coverage probabilities.}
    \begin{tabular}{ |p{1.25cm}|p{2cm}|p{2.5cm}|p{1.5cm}|p{2cm}|p{0.75cm}| p{0.75cm}| p{1.5cm}|}

\hline
 & Reference point & Normal data (size=1500) & Normal data & Observation noise data &  M1 & M2 & Average \\
\hline
MC & $\mu$ & 0.74 &  0.66 & 0.36 & 0.68& 0.64 & 0.61\\
 & $\sigma$ & 0.73 & 0.67 & 0.65 & 0.71 & 0.57 & 0.66\\
\hline
MCMC & $\mu$ &0.81 & 0.69 & 0.37 & 0.67& 0.73 & 0.65\\
 & $\sigma$ & 0.86 &0.61 & 0.68 & 0.53 & 0.42 & 0.62\\
\hline
\end{tabular}

\end{table*}

\subsubsection{\textbf{Recommendations}}
The MCMC is considered the most reliable and complex method and is getting wider adoption in the last years~\cite{hilborn2003state}. We support this trend with the findings from the comparison in this paper, claiming that it can lead to more accurate results. However, the method must have enough data available. The MCMC also has the most significant potential for improvements in manufacturers data, as a more appropriate probability distribution can be proposed. On the other hand, MC has proved to be just as suitable as MCMC for these real-world data with fewer data samples and is more than fifty times faster than MCMC. Moreover, MC is designed to the use of cumulative distribution for any probability distribution. More details can also be found in \cite{Kostka}.


\section{Conclusions}
In this paper, we have proposed a workflow to conduct comparison of Monte Carlo methods for Industry 4.0 applications in terms of process KPIs. To validate the proposed workflow, we have compared MC and MCMC methods with real-world dataset from Smart Manufacturing. The experimental results have shown that the two Monte Carlo methods can be selected based on the data availability. When there is enough data, MCMC is recommended in the Smart Manufacturing application due to its reliability and potential of improvement, whereas under the data sparsity, MC is recommended due to minor computational complexity. Thus, the Monte Carlo method selection in Industry 4.0 depends on the context such as data availability, target KPI, knowledge of the underlying data distribution. As future works, we plan to look into improvements of MCMC by evaluating additional distributions (e.g, exponential distribution), introducing more sampling methods (e.g., Gibbs sampling method), and considering disaggregation of Smart Manufacturing data.

\textbf{Acknowledgment.} The research was supported from ERDF/ESF "CyberSecurity, CyberCrime and Critical Information Infrastructures Center of Excellence" (No. CZ.02.1.01/0.0/0.0/16\_019/0000822).

\input{Main.bbl}


\end{document}

%% file: Main.bbl

%% file: Main.bbl
\begin{thebibliography}{10}
\bibitem{hilborn2003state}
R.~Hilborn, ``The state of the art in stock assessment: where we are and where
  we are going,'' \emph{Scientia Marina}, vol.~67, no.~S1, pp. 15--20, 2003.
\end{thebibliography}
